\documentclass[10pt,conference, a4paper]{IEEEtran}
\usepackage[utf8]{inputenc}
\usepackage[T1]{fontenc}
\usepackage[usenames, dvipsnames]{color}
\usepackage{cite}

\ifCLASSINFOpdf
  \usepackage[pdftex,final]{graphicx}
  \graphicspath{{../pdf/}{../jpeg/}}
  \DeclareGraphicsExtensions{.pdf,.jpeg,.png}
\else
  \usepackage[dvips]{graphicx}
  \graphicspath{{../eps/}}
  \DeclareGraphicsExtensions{.eps}
\fi

\usepackage{tabularx}
\usepackage{caption}
\usepackage{subcaption}
\usepackage{algorithmic}
\usepackage{algorithm}

\usepackage[bookmarks=false, colorlinks=true, urlcolor=blue, linkcolor=black, citecolor=black]{hyperref}
\usepackage{hhline}
\usepackage{booktabs} 
\usepackage{fancyvrb}
\usepackage{amsmath}
\usepackage{amssymb}

\begin{document}
%
\title{One-to-One Matching of RTT and Path Changes}

\author{
\IEEEauthorblockN{Wenqin Shao, Jean-Louis Rougier \\and Antoine Paris}
\IEEEauthorblockA{Telecom ParisTech\\
{\footnotesize \{firstname.lastname\}@telecom-paristech.fr}}
\and
\IEEEauthorblockN{Fran\c{c}ois Devienne \\and Mateusz Viste}
\IEEEauthorblockA{Border 6\\
{\footnotesize  \{firstname.lastname\}@border6.com}}
}

\maketitle
\pagestyle{empty}

\begin{abstract}
Route selection based on performance measurements is an essential task in inter-domain Traffic Engineering. 
It can benefit from the detection of significant changes in RTT measurements and the understanding on potential causes of change.
Among the extensive works on change detection methods and their applications in various domains, few focus on RTT measurements. It is thus unclear which approach works the best on such data.  

In this paper, we present an evaluation framework for change detection on RTT times series, consisting of:  
1) a carefully labelled 34,008-hour RTT dataset as ground truth;
2) a scoring method specifically tailored for RTT measurements.
Furthermore, we proposed a data transformation that improves the detection performance of existing methods.

Path changes are as well attended to. We fix shortcomings of previous works by distinguishing path changes due to routing protocols (IGP and BGP) from those caused by load balancing. 

Finally, we apply our change detection methods to a large set of measurements from RIPE Atlas. 
The characteristics of both RTT and path changes are analyzed; the correlation between the two are also illustrated. 
We identify extremely frequent AS path changes yet with few consequences on RTT, which has not been reported before.

\end{abstract}


\IEEEpeerreviewmaketitle

\section{Introduction}
\label{sec:intro}
Border Gateway Protocol (BGP) route selection
is agnostic of transmission performance, such as Round-Trip Time (RTT).
Supplementary Traffic Engineering (TE) scheme is thus needed.
Route selection based on latency measurements~\cite{Akella2008, Shao2015b}
sends out traffic on paths with the smallest recent RTTs for each destination prefix.
Re-routing could be overwhelmingly frequent due to the the noisy nature of RTT measurements, e.g. based on last RTT measurement~\cite{Shao2015b}. 
Meanwhile, too much smoothing could delay the reaction to sudden RTT changes.
In both cases, the most appropriate setting may vary from path to path, and is hence ad hoc.
We therefore advocate route selection based on the detection of significant changes in RTT measurements.
To that end, we study the techniques of change detection and how detected RTT changes correlate to network events.

Path changes and congestion are known to be the major reasons for RTT changes.
It is generally agreed that inter-domain routing changes impact the RTT level greatly.
Pucha et al.~\cite{Pucha2007} showed that inter-domain routing changes cause larger median RTT variation than intra-domain ones.
Rimondini et al.~\cite{Rimondini2014} confirmed that $72.5\%$ BGP route changes in their study are associated with RTT change.
Similar observations were made in a large Content Delivery Network (CDN), where inter-domain routing changes are responsible for more than $40\%$ of severe user experience degradation~\cite{Zhu2012}.

Intra-domain events are no less important. Pucha et al.~\cite{Pucha2007} discovered that intra-domain path changes can cause RTT changes of comparable amplitude as inter-domain ones.
Moreover, they pointed out that it is intra-domain path changes, not congestion, that are responsible for the majority ($86\%$) of RTT changes. 
A different claim was however made by Schwartz et al.~\cite{Schwartz2010}. They found out that most RTT variation is rather within paths (i.e. due to congestion) than among paths (i.e. due to path changes).

Conflicts in previous works could be caused by the difference in locations from where measurements were launched.
For instance, Chandrasekaran et al.~\cite{Chandrasekaran} observed that AS path changes only have marginal impact on RTT in the core of Internet, while previous works~\cite{Pucha2007, Schwartz2010} include as well access networks.
Results might as well change over time. For instance, the ``flattened'' Internet topology, the increasing amount of traffic in private CDN over the last decade~\cite{Labovitz2011, Roughan2011} might have changed the characteristics of path change and congestion, and consequently how they impact RTT.

Bearing this in mind, we emphasize the efforts on methods and tools enabling iterative
analysis on the relationship between RTT and path changes over time, rather than one shot observation or analysis on a specific dataset.

The discussion and discovery of previous works are enlightening, yet their methods of processing RTT measurements can hardly fuel RTT change triggered intra-domain TE.
In \cite{Pucha2007, Schwartz2010, Chandrasekaran},
RTT measurements are first grouped by underlying paths; impact of path changes are then estimated through comparison of associated RTT statistics, e.g. percentiles.
In our TE scheme, RTT rather than path are more intensively measured~\cite{shao2016}.
This is because path measurements do not necessarily reflect all the changes in RTT, meanwhile being more resource-consuming.
For example, congestion in upstream network can only be learnt through change detection on RTT measurements.

Among the extensive studies on change detection methods and their applications in various  domains~\cite{Chen2001,Haynes2016},
Rimondini et al.~\cite{Rimondini2014} are among the first to employ change detection in network measurements analysis.
However, they tuned the detection sensitivity in a way that detected changes correlate best to the BGP changes of the destination prefix, which potentially ignores the RTT change due to intra-domain changes and congestion.
Plus, such tuning is potentially required for each individual RTT time series, thus hard to scale.
To achieve more general approach decoupled from path measurements, we propose an evaluation framework for the selection and calibration of change detection methods for RTT measurements.

IP-level load balancing (LB) is few discussed in previous investigations on the relationship between RTT and path changes.
Schwartz et al.~\cite{Schwartz2010} regarded all paths between a source-destination pair as ``parallel paths'' and found out that RTTs over these paths were mostly overlapping.
However, there are two kinds of transitions among ``parallel paths'' that need to be distinguished.
They are 1) IP path changes caused by protocol level route recalculation
and 2) those caused by LB mechanisms. 
Intra-domain path changes before the era of LB haven been shown to be responsible for important RTT changes~\cite{Pucha2007}. 
On the other hand, LB paths are of equal/close administrative cost, hence similar characteristics~\cite{Augustin2011}. 

This work proposes a set of methods detecting and correlating individual RTT and path changes, preparatory to RTT change triggered inter-domian route selection.
The pursuit is thus not to answer again which network event has most significant impact on RTT.
The contribution brought forth are: 1) a customized scoring scheme together with a carefully labelled dataset are presented as evaluation framework; 2) a simple data transformation is proposed and shown to improve the detection performance of existing methods; 3) a heuristic is developed to distinguish path changes caused by routing change from those by LB.
We further uncovered the characteristics, e.g. change detection sensitivity,  and remaining issues with the proposed methods.

The remainder of the paper is organized as follows. Sec.~\ref{sec:data} points to the code repository of this work and provides details on the data collection. Sec.~\ref{sec:rtt} first offers a primer on change detection methods; then presents our propositions on change detection method evaluation. Sec~\ref{sec:path} reveals and addresses the challenge of distinguishing routing changes from LB path changes for RIPE Atlas built-in measurements. Sec.~\ref{sec:corr} correlates the detected RTT and path changes.


\section{Code space and data}
\label{sec:data}
The main code space of this work is made public on Github with documentation: \url{https://github.com/WenqinSHAO/rtt}.
The implementations of proposed methods are decoupled from the context of this project, and thus can easily be employed elsewhere.  

We applied our methods on RIPE Atlas built-in measurements~\cite{atlas} and performed data analysis.
These measurements are openly available so that the results of this work can be reproduced by other researchers or compared to alternative approaches.
We collected RIPE Atlas built-in ping and traceroute measurement toward DNS b-root (measurement ID 1010, 5010) from 6029 v3 probes located in 2050 different ASes, 153 countries from 2016-10-01 to 2017-01-01~\footnote{Measurements to other destinations might as well do. The fact whether the destination is anycast or not is of few importance in this work. The focus is on method rather than on a specific dataset.}.
184,358,516 ping and 23,507,910 traceroute measurements are collected and analyzed.
The traceroute measurements flowed through 3036 ASes, 120 IXPs, containing 10720 different AS paths.

\section{RTT change detection}
\label{sec:rtt}
RTT traces, like many other time series, may undergo sudden changes in level or volatility,
generally caused by path change or congestion.

The moments that cut a time series into segments of different characteristics are called \textit{changepoints}.
The problem of detecting the most appropriated changepoints is known as changepoint detection.
Which method (among the wide variety of existing ones) is the most appropriate for Internet RTT time series is still not stated. Moreover, many changepoint detection methods are parametric. Identifying the best settings for these methods remains challenging.
One fundamental issue in addressing the above problems is the lack of an evaluation framework.

In this section, we first introduce changepoint detection method.
We explain the parameters to be set and their impacts on the detection results.
Then, we dissect the challenges in building an evaluation framework and describe our attempt in solving them.
Finally, we choose several state-of-art changepoint detection methods and evaluate their performance with the proposed evaluation framework.

\subsection{Changepoint detection}
\label{sec:cpt}
One common approach translates the quest of finding the best changepoints into the following optimization problem \footnote{Other formulations exist. A wider literature can be found in \cite{Haynes2016, Eckley2011}. We focus on this approach in this work since it has well maintained libraries that prevent potential issues regarding the implementation~\cite{Killick2013a, Haynes2016}.}.
Assume we are given a sequence of data, $y_{1:n} = (y_1, y_2,...y_n)$.
We expect changepoint detection method to produce $m$ ordered changepoints, $\tau_{1:m} = (\tau_1, \tau_2,...\tau_m)$.
$\tau_i$ is the position of $i^{th}$ changepoints and takes value from in ${1,..,n-1}$.
We define $\tau_0 = 0$ and $\tau_{m+1} = n$.
Together with the detected $m$ changepoints, they cut $y_{1:n}$ into $m+1$ segments, with the $i_{th}$ segment containing $y_{\tau_{i-1}+1:\tau_i}$.
For each segment, a cost is calculated. The detection method seeks to minimize the cost sum of all the segments: $\sum_{i=1}^{m+1}[C(y_{\tau_{i-1}:\tau_i-1})] + \beta f(m).$
Here $C$ is a cost function while $\beta f(m)$ is a penalty to prevent over-fitting ---  the two major parameters to be set.

One commonly used cost function is minus of the maximum log-likelihood of the segment following a certain distribution~\cite{Killick2011,Horvath1993,Chen2001}:
$C(y_{s:t}) = - \max_\theta \sum_{i=s}^t \log f(y_i|\theta)$.
Here $f(y|\theta)$ is a density function with distribution parameter $\theta$. 
In such case, the choice of cost function is restrained to the choice of distribution types.
Currently supported distributions in \cite{Killick2013a} are: Normal, Exponential, Gamma and Poisson.
A recent progress proposes a cost function based on empirical distribution likelihood, where the specification on distribution type is not necessary. It is thus a non-parametric method~\cite{Haynes2016}. 

When it comes to penalty, $f(m)$ is generally a function linear to the number of parameters introduced by $m$ changepoints: 
$m + (m+1)dim(\theta)$ \footnote{$dim(\theta)$ is the dimension of $\theta$. In the case of Normal distribution, $dim(\theta) = 2$.}.
Common choices of $\beta$ are information criteria, such as Akaike’s Information Criterion (AIC) with $\beta=2$, Schwarz Information Criterion (SIC, also known as BIC) with $\beta=\log n$, Hannan-Quinn Information Criterion with $\beta = 2 \log \log n$, and Modified BIC (MBIC) with 
$\beta f(m) = -\frac{1}{2} [3f(m)\log n + \sum_{i=1}^{m+1} log(r_i - r_{i-1})]$, where $r_i = \tau_i/n$.
We have MBIC $>$ BIC $>$ Hannan Quinn. Note that larger penalty value leads to less sensitive detection.

\subsection{Evaluation framework}
\label{sec:frame}

An evaluation framework should be composed of two parts: 1) datasets of ``ground truth'', 2) a scoring method.
We are not aware of any RTT time series labelled with moments of change that are publicly available as of this writing.
We manually labelled 50 real RTT time series from RIPE Atlas containing 408,087 RTT measurements.
Details are given in Sec.~\ref{sec:label}.

As for the scoring method, classic true/false positive classification is too rigid for 
both manual labelling and change point detection, see Sec.~\ref{sec:score}.
We argue that a slight shift in time could be tolerated.
We propose weighting each actual RTT changes according to their operational importance. 

\subsection{Scoring method}
\label{sec:score}
We assume ground truth $T_{1:k}$ containing $k$ positions in $y_{1:n}$ indicating moments of actual change, while $\tau_{1:m}$ is the output of changepoint detection.
A classic True Positive ($TP$) is a $\tau_j, \exists T_i \in T_{1:k}, T_i = \tau_j$, a False Positive ($FP$) otherwise. False Negative ($FN$) composes of $\{T_i \mid \nexists \tau_j \in \tau_{1:m}, \tau_j = T_i \}$.
However, there are many times during labelling that finding a clear cut position for certain RTT changes is difficult for human beings, and the labelled moment of change might reasonably vary within a range.
It is therefore too harsh to require exact match from change detection.

Introducing a window of tolerance $w$, we are confronted with an issue where 
$T_i$ can be detected by multiple detection outputs $ \{\tau_j \mid |\tau_j - T_i| \leq w \}$. 
Symmetrically, $\tau_j$ can be associated with several $T_i$ within the tolerance window.
If many-to-many mapping between $T_{1:k}$ and $\tau_{1:m}$ is allowed, the number of $TP$ could be overestimated. while $FN$ underestimated.

One straightforward solution adopted by Numenta is that a $T_i$ can only be detected
by the closest detection in the window $\hat \tau_i$~\cite{Lavin2016}. All the rest detected changes will be ignored.
The problem with this approach is that the actual change that is closest to the above $\hat \tau_i$, denoted as $\hat T_i$, doesn't necessary satisfy $\hat T_i = T_i$. 
Such discrepancy indicates that the mapping between ground fact and detection is potentially not optimal in two aspects 1) $\hat T_i$ could end up undetected even though there are detection points within the window; 2) the overall time shift between truth and detection is not necessary the minimum.

We henceforth define an optimal mapping between $T_{i:k}$ and $\tau_{i:m}$ with shift tolerance $w$ , $MP = \{(T_x, \tau_y)\} \mid |T_x - \tau_y| \leq w \}$, as one that first maximizes $|MP|$, the size of $MP$, and then minimizes the total shift $\sum_{(T_x, \tau_y) \in MP} |T_x - \tau_y|$. 
We first construct a bipartite graph with cost $G = (V \cup W, E)$.
$V \cup W$ are the vertices, where $V = T_{i:k}$ and $W = \tau_{1:m}$.
Edge $E$ is composed of all ground truth and detected change point pairs that are within the window, $E = \{(p, q) \mid |p - q| \leq w, p \in V, q \in W \}$.
The cost of each edge is defined as the distance/shift in time between ground truth and the detected change point $C(e) = |p-q|, e \in E$.
The problem of finding the optimal mapping $MP$ is translated into finding the \textit{minimum cost maximum-cardinality matching} of $G$, for which the Hungarian algorithm is known as the best option.

For each detection $\tau_j$, if $\exists m \in MP, \tau_j \in m$, it is regarded as a $TP$, otherwise as a $FP$.
All the ground truth without matched detection $\{T_i \mid \nexists m \in MP, T_j \in m\}$ contributes to $FN$.
$Precision$ of the changepoint detection method, defined as  $\frac{TP}{TP+FP}$, can be interpreted as the fraction of detection that is relevant or useful.
$Recall$, defined as $\frac{TP}{TP+FN}$, can be regarded as the fraction of all ground truth change points that the method can successfully detect.

As mentioned early, not all RTT changes are equally important. 
We propose to weight a ground truth change $T_i$ according to the following three elements: 
1) the length of RTT segment following $T_i$, i.e. $T_{i+1} - T_i$;
2) the RTT level difference across $T_i$, denoted as $M_i$; 
3) the RTT volatility difference across $T_i$, denoted as $\Delta_i$.
More formally for each $T_i \in T_{i:k}$, with $T_0=1, T_{k+1} = n$, we define 
$M_i = |Median(y_{T_{i-1}+1:T_i}) - Median(y_{T_i+1:T_{i+1}})|.$
We use median instead of mean in the purpose of reducing the impact of abnormally large RTT measurements.
We define $\Delta_i = |Std(y_{T_{i-1}+1:T_i}) - Std(y_{T_i+1:T_{i+1}})|.$
We define empirically the weight associated to each $T_i \in T_{1:k}$ as: $\Omega_i = MAX(\log_2\frac{T_{i+1} - T_i}{\rho}, 0) \times (M_i + \Delta_i).$
Here $\rho$ is a threshold for RTT segment length. 
If $T_i$ leading to an RTT segment shorter than $\rho$, we ignore it in calculating $Recall$.
The intuition behind this weighting is that RTT changes of large level or volatility are in practice regarded as more important. $\rho$ and $w$ tolerance window are set to $8min$ in this work, corresponding to two ping measurement intervals. 

We can henceforth formulate a `weighted' version of the $Recall$ metric to better reflect the operational importance of detected RTT changes: $Recall_W = \frac{\sum_{i, T_i \in TP} \Omega_i}{\sum_{j=1}^k \Omega_j}.$

We use the $F_2$ score to consolidate $precision$ and $recall$, where recall is weighted twice as important as precision: $F_2 = (1+2^2) \times \frac{Precision \times Recall_W}{2^2Precsion + Recall_W}.$
The practical implication of this choice is that handling some FPs is less unwanted than missing out some important RTT changes.

\subsection{Labelling changes in RTT time series}
\label{sec:label}

\begin{figure}[!htb]
\centering
\includegraphics[width=.48\textwidth]{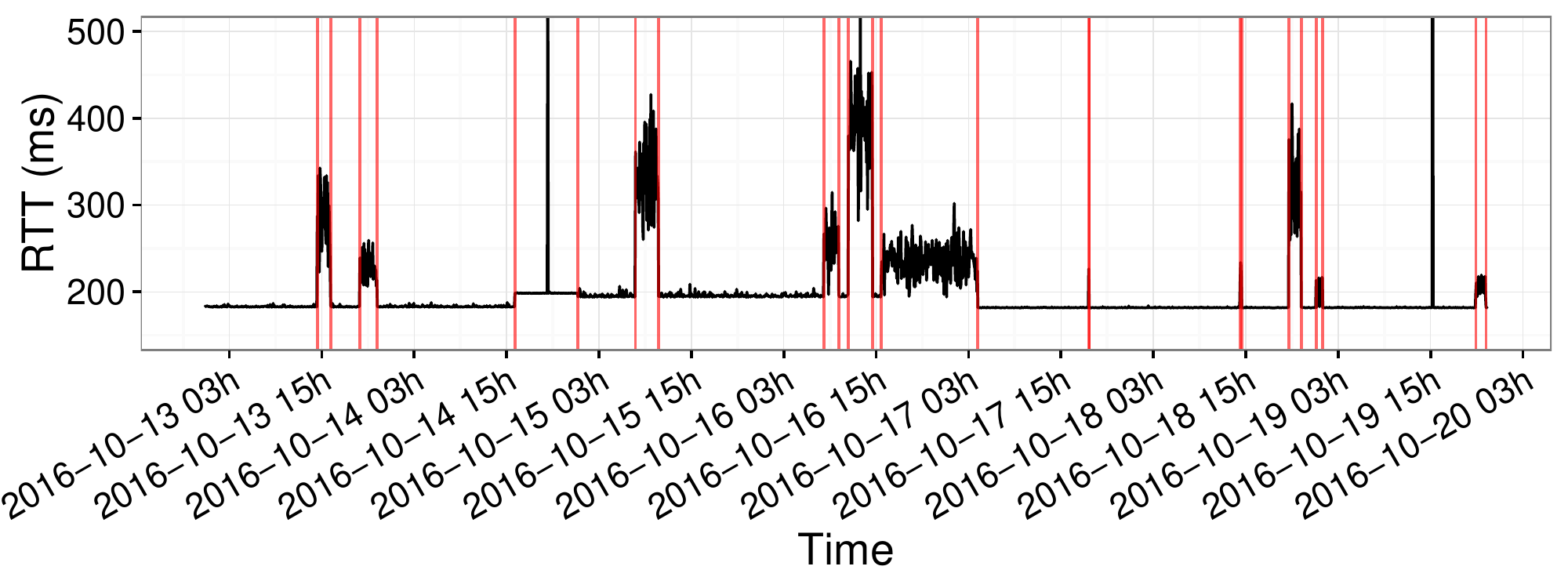}
\caption{First 2500 
Datapoints of an artificial RTT time series (one datapoint every 4min). 
Red vertical lines correspond to generated changes.}
\label{fig:art_example}
\end{figure}

In order to determine which detection method works the best on RTT time series, 
a dataset with \textit{a priori} labelled moments of RTT change is required, serving as ground truth, in the above presented scoring method. It's quality is essential to the relevance of evaluation results.

There are two approaches to a labelled ground truth: 1) artificially generated data; 2) real data with labels.
Real data is naturally the preferred choice and 
can only be labelled by humans with domain knowledge in absence of systematic automation, which is however tedious and error-prone. 
Therefore, it is of importance to first design tools facilitating the labelling and second to evaluate the quality of so produced `ground truth'.
More specifically, 1) a set of tools for interactive visual inspection (for RTT time series and labels) are developed to minimize human errors~\footnote{\url{https://github.com/WenqinSHAO/rtt_visual.git}}; 2) we fabricated a synthetic RTT dataset with known moments of actual change, and compared the human detection results to generated change moments~\footnote{\url{https://github.com/WenqinSHAO/rtt_gen.git}}. The labellers are the authors of this work, who are researchers/graduate students in networking. 

The synthetic dataset contains 20 RTT timeseries 
representing 8646 hours of RTT measurements with 935 generated changepoints.
An example of these synthetic RTT trace is shown in Fig.~\ref{fig:art_example}.
Each trace contains several stages of random RTT level representing different underlying paths.
Each path has its own Markov process deciding the chance of getting into/out of a congestion phase. 

\begin{figure}[!htb]
\centering
\includegraphics[width=.36\textwidth]{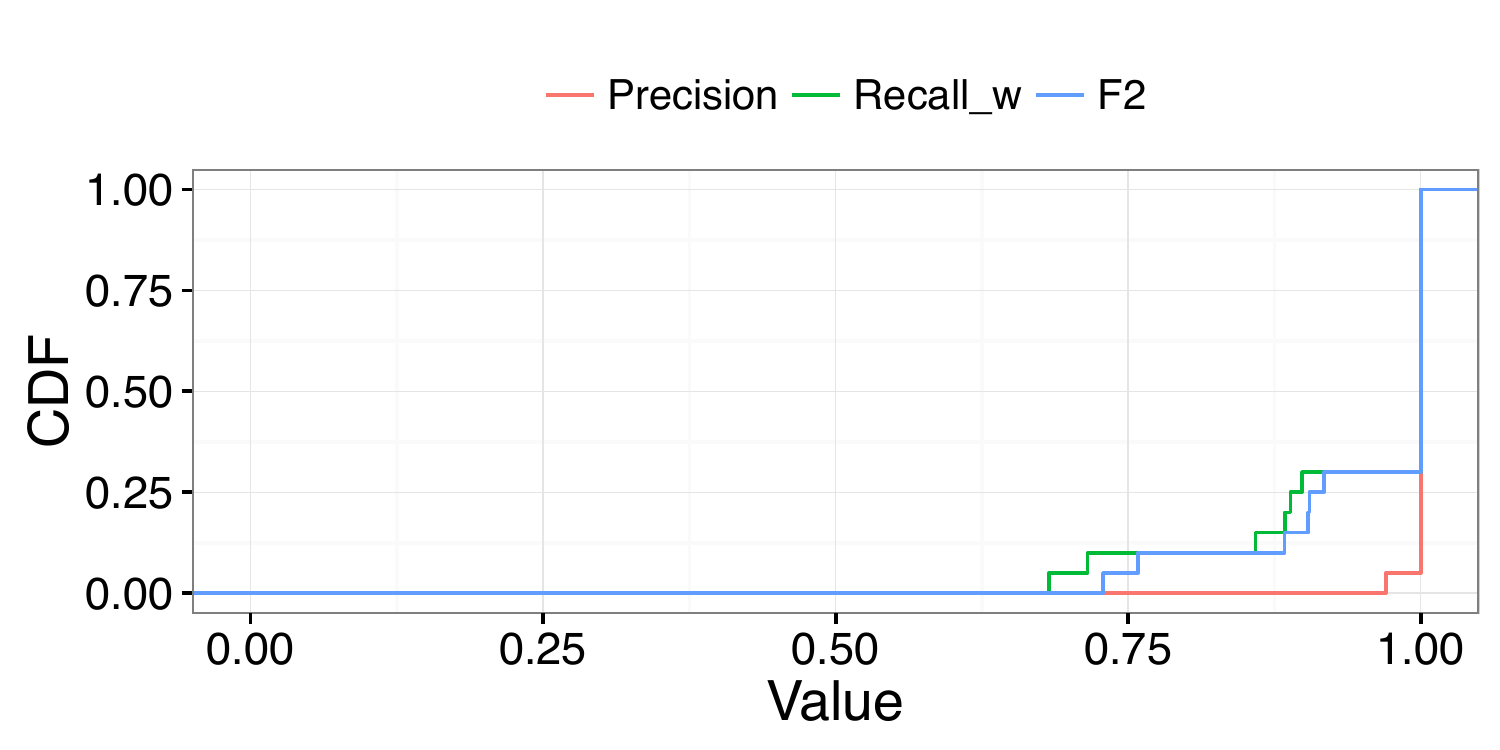}
\caption{$Presicion$, $Recall_W$ and weighted $F_2$ of human labellers on synthetic dataset.}
\label{fig:antoine_eval}
\end{figure}
The detection performance of human labellers on the synthetic dataset is shown in Fig.~\ref{fig:antoine_eval}~\footnote{Note that the labellers had no idea these series were synthetic, since they are mixed up with real traces during labelling.}. 
Human labellers have $100\%$ for both $Precision$ and $Recall$ on 14 traces.
For the rest, the $Precision$ remains high. A few changes are miss out, but their total weight remain limited.

Real RTT traces of various characters are selected from RIPE Atlas to construct the ground truth dataset. Some are full of fluctuations; some contain periodic congestion, some have many stage changes, etc.
The entire dataset represents more than 34,008 hours, i.e. 1417 days, of RTT measurements. 1047 changepoints were identified by the labellers.
The labelled RTT traces along with the synthetic traces are all available in the main project repository given in Sec.~\ref{sec:data}.

\begin{figure*}[!ht]
\centering
\includegraphics[width=.9\textwidth]{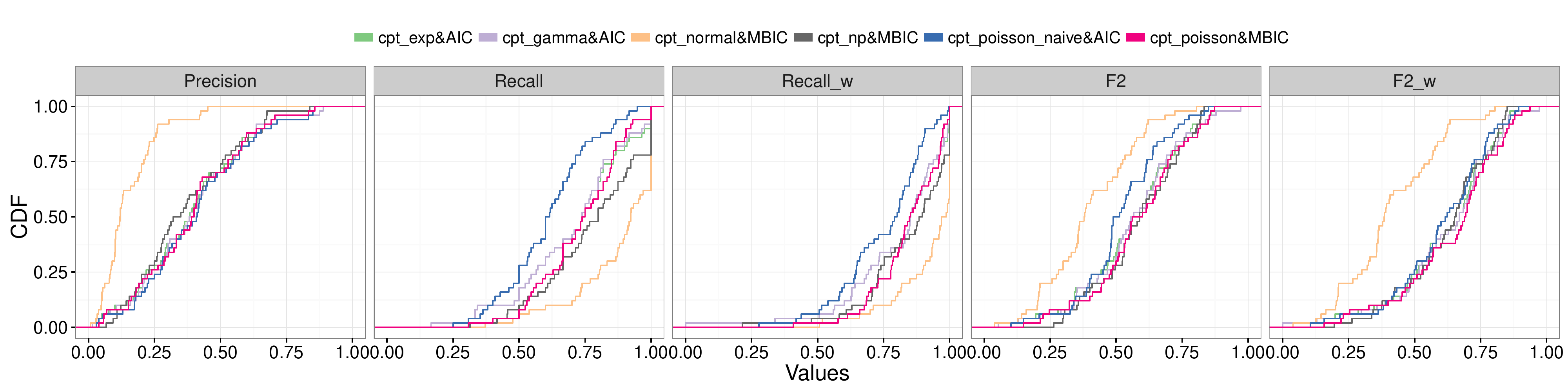}
\caption{$Precision$, $Recall$, $Recall_W$, $F_2$ and $F_2W$ with weighted recall on real RTT traces.}
\label{fig:real_eval}
\end{figure*}

\subsection{Candidate changepoint methods}
\label{sec:method}
According to the primer on changepoint detection in Sec.~\ref{sec:cpt}, there are two major parameters for a changepoint method formulated in that way: penalty and cost function/distribution.

We consider all the information criterion introduced (AIC, BIC, MBIC and Hannan-Quinn), and all the supported distribution types, including the non-parametric approach based on empirical distribution.

With some preliminary tests, we quickly realized that detection with Normal distribution tend to be over-sensitive, while Poisson, Exponential distribution are too numb.
It is because the mean and variance of Normal distribution are independently controlled by two parameters, which increases the chance of fitting subtle changes either in level or volatility.
Meanwhile, the mean and variance of Poisson and Exponential distribution are coupled by one parameter,
which restrains their freedom of adjustment~\footnote{Poisson, mean=variance=$\lambda$; Exponential, mean/variance=$\lambda$, mean=$1/\lambda$.}~\footnote{Gamma, mean/variance=$\beta$, mean=$\alpha/\beta$. \cite{Killick2013a} requires \textit{a priori} input for $\alpha$, which decides overall sensitivity. Therefore, only $\beta$ is tuned in finding changepoints.
We tried $\alpha$ from 1 to 100 on the labelled dataset. None of them outperforms the best setting shown later on. Due to space limit, we no longer consider Gamma distributions.}.
For instance, for a path including trans-Pacific links, we shall expect a minimum RTT above 80ms, in which case the corresponding Poisson distribution could easily tolerate several RTT deviations of 20ms, which is already non-negligible.


To boost the detection sensitivity with Poisson and Exponential distribution, we propose a \textit{data transformation}: subtracting the RTT time series by its minimum value (baseline) to lower down its overall RTT level~\footnote{Note that timeout measurements are set to 1000ms. For Poisson distributions, RTT values are rounded to the closest integer.}. 
Changes are then detected for the baseline-removed RTT time series when assuming Poisson and Exponential distribution.
Such setting is denoted as \texttt{cpt\_poisson} and \texttt{cpt\_exp} respectively.
For the sake of comparison, we also consider Poisson distribution without data transformation and denote it as \texttt{cpt\_poisson\_naive.}
Normal distribution and non-parametric approach are applied directly on initial RTT measurements.
They are denoted as \texttt{cpt\_normal} and \texttt{cpt\_np} accordingly.

\subsection{Evaluation of changepoint methods}
\label{sec:eval}
Before evaluating with the scores defined in Sec.~\ref{sec:score}, one might wonder whether the RTT segments labelled by human beings already (Sec.~\ref{sec:label}) follow principally a specific distribution, and whether that distribution leads to the best detection performance.
We performed distribution test for 813 RTT segments longer than 20 datapoints against each of the discussed distribution types under corresponding data transformation (Sec.~\ref{sec:method}). 71 follow Normal distribution, 13 follow Poisson distribution, 
11 follow Exponential distribution~\footnote{Significance level 0.05. Shapiro-Wilk Normality test; 
Chi-squared test for Poisson; Kolmogorov-Smirnov test for Exponential. Distribution parameters are estimated through Maximum Likelihood Estimation.}. None of these distributions seems to to have dominant popularity among the labelled RTT segments.

Changes are detected for selected real RTT traces with all distribution types.
The detection performance in terms of $Precision$, $Recall$, $Recall_W$, $F_2$ and weighted $F_2$ under optimal penalty are given in Fig.~\ref{fig:real_eval}.
More than $75\%$ of changes in terms of weight can be detected for more than half of the traces with any distribution.
All distribution types have better score in terms of recall and $F_2$ with their weighted variation, indicating some changepoints missed out are indeed of little operational importance.
Normal distribution has more fitting RTT segments than others, however its overall $F_2$ score (weighted or not) is not outstanding. This suggests that the goodness of fit is not a guarantee for detection performance.

Fig.~\ref{fig:real_eval} confirms that the detection with Normal distribution is over-sensitive even with MBIC, the largest adaptive penalty setting. For the other methods, their performances are rather close. 
\texttt{cpt\_poisson} seems to have a slight advantage according to weighted $F_2$.
Compared to \texttt{cpt\_poisson\_naive}, \texttt{cpt\_poisson} achieves higher $Recall_W$ without obviously sacrificing $Precision$.
As a matter of fact, without data transformation, assuming Exponential distribution detects no changepoint for a big part traces in the real RTT dataset.
These imply that the proposed data transformation has the potential to improve detection performance for these distributions. 

\subsection{Detecting changes for collected ping measurements}
\label{sec:cpt_trace}
\texttt{cpt\_poisson} and \texttt{cpt\_np} with MBIC are used to detected RTT changes for all the 6029 collected ping measurements.
We consider \texttt{cpt\_poisson} as it is the best performing one, though by a small margin.
\texttt{cpt\_np} is included as it performs well and its cost function follows a different principle.


\begin{figure}[!htb]
    \centering
    \begin{subfigure}[b]{.24\textwidth}
	\centering
	\includegraphics[width=\textwidth]{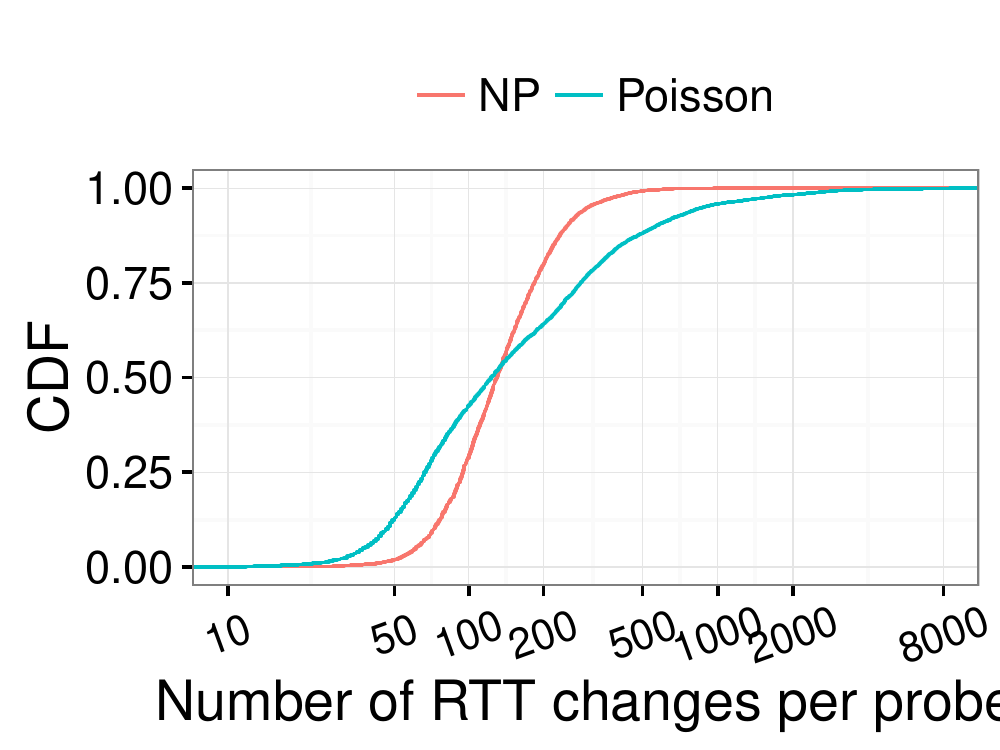}
	\caption{\footnotesize CDF.}
	\label{fig:rtt_ch_count_cdf_cmp}
	\end{subfigure}
	\begin{subfigure}[b]{.24\textwidth}
	\centering
	\includegraphics[width=\textwidth]{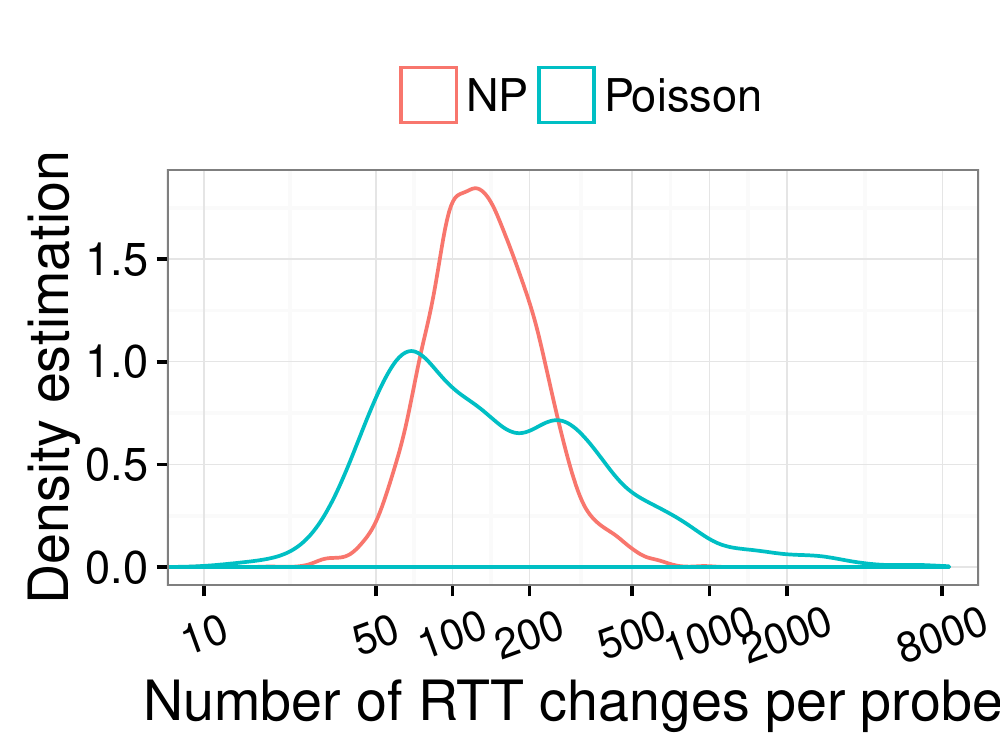}
	\caption{\footnotesize Density.}
	\label{fig:rtt_ch_count_density_cmp}
	\end{subfigure}
\caption{RTT changepoints number distribution with different detection methods under MBIC.}
\label{fig:rtt_ch_count_cmp}
\end{figure}
Fig.~\ref{fig:rtt_ch_count_cmp} shows the distribution of RTT change numbers per probe. 4844 probe traces each containing more than 30,000 ping measurements are considered. 
854,626 RTT change are detected by \texttt{cpt\_np}.
\texttt{cpt\_poisson} almost doubled this number with 1,638,858 RTT changes.
However, the median change numbers for both methods is however the same (122).
Fig.~\ref{fig:rtt_ch_count_density_cmp} shows that the change number by \texttt{cpt\_poisson} spreads over a much wider range.
With \texttt{cpt\_poisson}, 711 probes ($11.86\%$) have more than 500 changes, while only 35 probes ($0.58\%$) with \texttt{cpt\_np} experienced that many changes.
This is probably because the cost function of \texttt{cpt\_np} bases on the estimation of quantiles (by default 10 quantiles used, more can be set) of empirical distribution. The dimension of $\theta$ is much larger than Poisson and Normal distribution. The penalty value increases hence much faster for \texttt{cpt\_np} when new changepoint is added, which prevents extremely large number of changepoints per probe trace.

\begin{figure}[!thb]
\centering
\includegraphics[width=.48\textwidth]{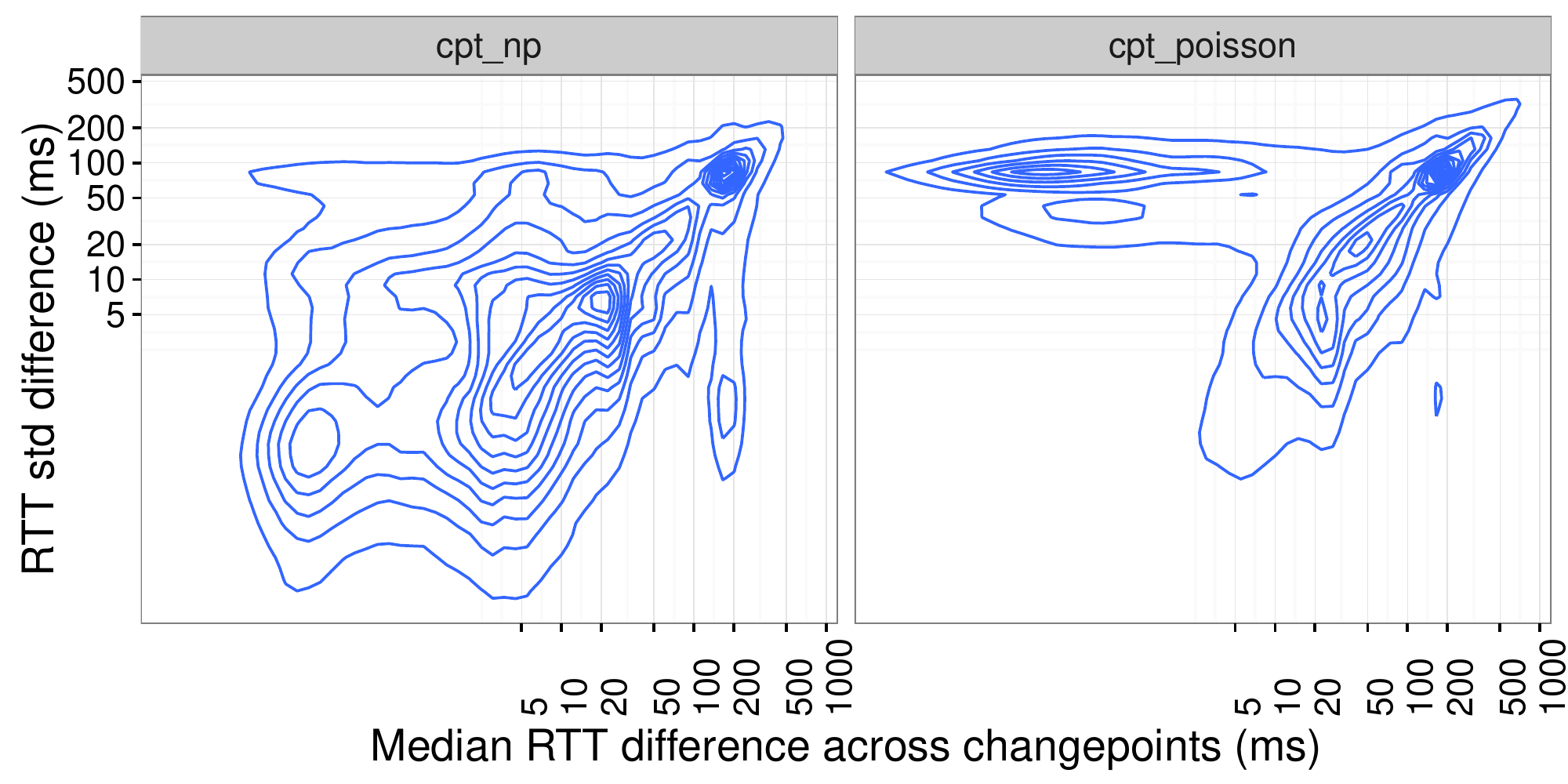}
\caption{Density estimation of RTT changepoints characteristics.}
\label{fig:rtt_chara_cmp}
\end{figure}
Fig.~\ref{fig:rtt_chara_cmp} compares the characters of all the RTT changes detected by the two methods.
Each RTT change is described by the associated level and volatility difference, i.e. $M$ and $\Delta$ defined in Sec.~\ref{sec:score}. 
Overall, RTT changes detected by \texttt{cpt\_poisson} are of larger $M$ and $\Delta$.
331,062 ($38.71\%$) changes with $M$ and $\Delta$ both smaller than 5ms are detected by \texttt{cpt\_np}.
With \texttt{cpt\_poisson}, there are only 87,021 ($5.31\%$) such changes of few importance in networking.
Meanwhile, a relatively bigger fraction of RTT changes detected by \texttt{cpt\_poisson} have large $\Delta$ but small $M$, more precisely 319,541 ($19.49\%$) with $M<5ms, \Delta>50ms$.
They are mostly associated with short RTT segments caused by frequent timeouts in certain probe traces. For example, probe 20854 had 2308 timeout measurements dispersed in the entire trace.
\texttt{cpt\_poisson} appears to be more sensitive to such sporadic but short living deviations.

\textbf{Wrap-up} In this part, we described an evaluation framework adapted to change detection on RTT time series.
A data transformation to improve the detection sensitivity is proposed,
with which, Poisson distribution + MBIC achieves the most appropriate balance between sensitivity and relevance. The nuance between \texttt{cpt\_np} and \texttt{cpt\_poisson} will be further discussed in Sec~\ref{sec:corr}.

\section{Path change detection}
\label{sec:path}
Trivial as it may sound, detecting IP path changes is challenging for RIPE Atlas built-in traceroute measurements.
The difficulties come from two aspects: 1) the wide deployment of IP-level Load Balancing (LB); 2) RIPE Atlas uses Paris traceroute with different Paris IDs every other measurement (incremented by 1, recycling between 0 and 15)~\cite{Augustin2006, Pelsser2013}.

IP paths taken by two neighbouring measurements can naturally differ -- 
load-balanced on different available paths.  
From this angle, plain IP path changes doesn't mean that there were topological or configuration changes that lead to any real routing change. 
On the other hand, having different Paris IDs every time can also be helpful in this context.  If traceroute were locked on a single Paris ID, it would then be impossible to detect routing changes that only affect paths corresponding to other Paris IDs.

\subsection{IP Forwarding Pattern change}
When a different IP path is measured with a same Paris ID,
there is potentially a routing change. We call this kind of IP path change an \textit{IP Forwarding Pattern} (IFP) change.
In the example below, the IFP change happens when Paris ID 2 begins to take IP path E instead of B. We refer to two measurements with same Paris ID but different IP paths as \textit{conflicting} measurements.
\begin{Verbatim}[fontsize=\footnotesize]
                             | IFP change
Paris ID: 0 1 2 3 4 .. 15 0 1|2 3 ..
IP Path:  A B B A A .. C  A B|E E ..
       A measurement series  | boundary -> forward
\end{Verbatim}

IFP changes can thus be identified by constructing a set of measurement series, each containing no conflicting measurements. Yet, across two series next to each other, there shall be as least one pair of conflicting measurements, otherwise they can be merged.
This can be done by moving the boundary of measurement series \textit{forward} to include non-conflicting measurements, till a conflict is encountered, as shown in the above example.
We call this approach \textit{forward inclusion}.

The drawback of \textit{forward inclusion} is that it potentially delays the detection of actual IFP changes.
This is because, when including non-conflicting measurements forwardly, a measurement series always has the chance to absorb measurements till it experiences all possible Paris IDs.
However an actual IFP change could happen before that moment.
An example of possibly delayed IFP change is given right below:
\begin{Verbatim}[fontsize=\footnotesize]
        !Possible position of actual IFP change        
.. 1|2 3!4 5 .. 15 0 1|2 3 4 5 ... 15 0 1 2 3 4 5 ..
.. B|B A!A C .. C  A B|E E A C ... C  A B E E A C ..
                      | IFP change forward inclusion
                      | backward <- boundary
\end{Verbatim}
With \textit{forward inclusion}, an IFP change will be detected at the 2nd appearance of Paris ID 2.
While the actual change probably happens at the 1st appearance of Paris ID 4, since starting from it, all the measurements are non-conflicting with the later measurement series.
The 1st appearance of Paris 2 and 3 are in fact a short deviation from a popular IFP.

Cases like this are highly possible, because networks tend to have some stable configurations that lead to a few dominant paths over time~\cite{Chandrasekaran, Pucha2007}. 
Deviations from dominant/popular IFPs are thus likely to be short living.
With RIPE Altas built-in measurements, they probably won't last long enough to experience all the Paris IDs~\footnote{It takes at least 450min (30min * 15) to go through all the 16 Paris IDs used in RIPE Atlas built-in traceroute.}.
To maximize the presence of popular IFPs, we only push backwardly the boundary obtained by \textit{forward inclusion} if 1) the latter measurement series is longer than the previous one; 2) the latter measurement series experiences all the Paris IDs at least twice.
We refer to this approach as \textit{backward extension}.
We show later on that IFP changes detected by \textit{backward extension} have a much larger chance matching with RTT changes.

\subsection{Detecting path changes}

AS-level path changes are as well detected after translating IP hops to ASN hops~\cite{routeviews}.
We didn't consider third-party address~\cite{Hyun2003, Zhang2010} and IP alias techniques\cite{Gunes2009,Keys2010a} in this operation.
It is because the focus is to detect changes instead of constructing an accurate Internet topology.
We did detect the presence of IXPs using the heuristics proposed by traIXroute~\cite{Nomikos2016}, since individual studies have shown that IXP could be involved in large RTT changes~\cite{kopp2016}.

\begin{figure}[!htb]
    \centering
    \begin{subfigure}[b]{.24\textwidth}
	\centering
	\includegraphics[width=\textwidth]{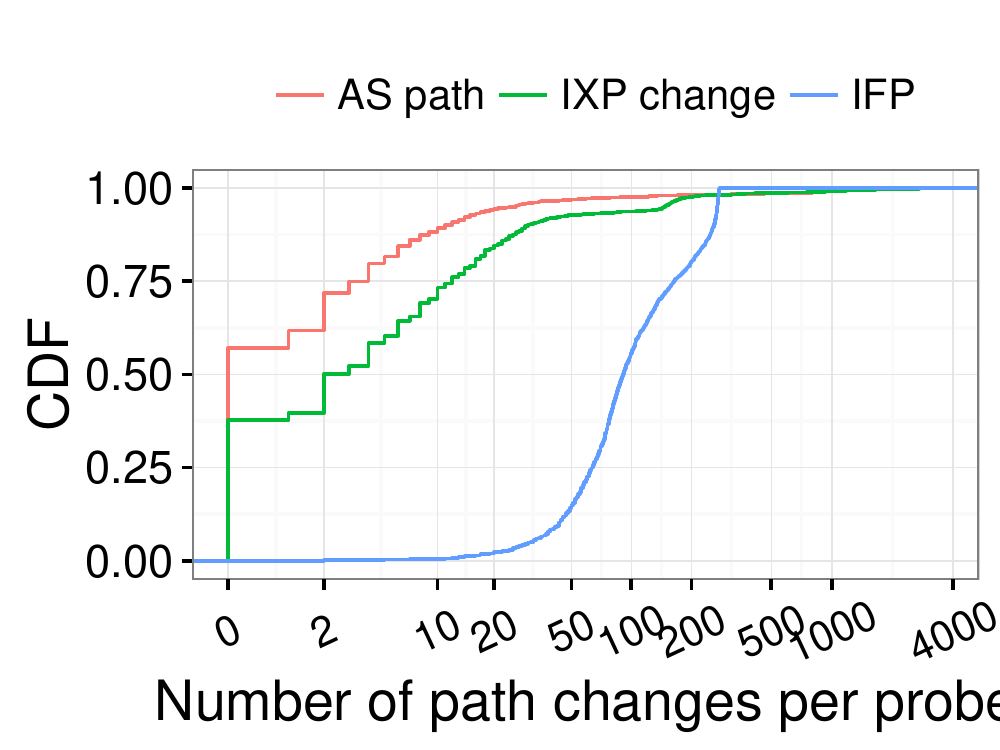}
	\caption{\footnotesize CDF.}
	\label{fig:path_ch_count_cdf_cmp}
	\end{subfigure}
	\begin{subfigure}[b]{.24\textwidth}
	\centering
	\includegraphics[width=\textwidth]{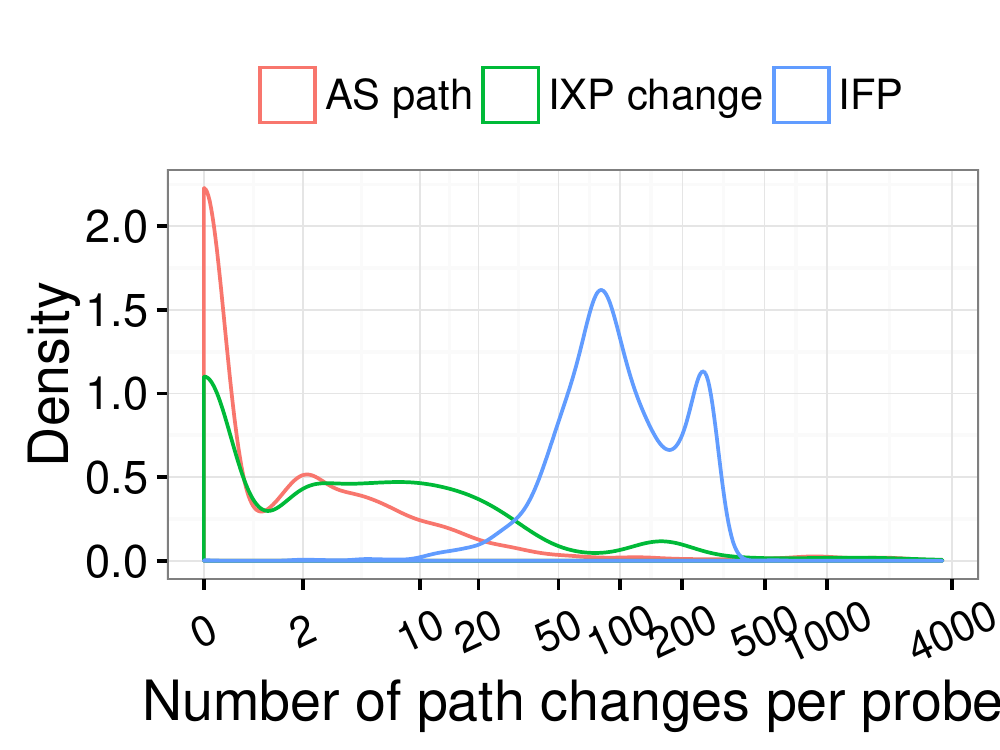}
	\caption{\footnotesize Density.}
	\label{fig:path_ch_count_density_cmp}
	\end{subfigure}
\caption{Path change times per probe trace distribution. One probe with most complete traceroute measurement is chosen for each AS. 2050 probes/ASes are inlcuded in the graph.}
\label{fig:path_ch_count_cmp}
\end{figure}
We consider only AS path changes where the difference starts from a hop position involving public ASNs in both AS paths.
Difference due to temporal presence of non-responding hops are ignored.
IXP change happens when the difference starts from a position involving at least one IXP hop in the two AS paths.
IFP changes, detected with \textit{backward extension}, not overlapped with AS path/IXP changes are considered.
They are potentially caused by intra-domain routing changes.

The distribution of number of path changes per probe trace
is illustrated in Fig.~\ref{fig:path_ch_count_cmp}.
One probe with the most complete traceroute measurements is selected for each of the 2050 source ASes.
1170 ($57.07\%$) of them experienced no AS path changes over the period of three months, indicating that the AS paths are in general very stable over time.
Still, 51 ($2.49\%$) probes underwent more than 100 AS path changes.
717 probes ($34.98\%$) didn't have any IXP change.
140 probes experienced frequent ($> 100$) IXP changes.
IFP changes are much more frequent than the other two path changes. 
Half of the selected probes experienced more than 90 IFP changes.
We investigate the nature of these path changes, together with their potential impact on RTT, in Sec.~\ref{sec:corr}.


\section{Correlation Study}
\label{sec:corr}
If a pair of RTT and path change are close in time, chances are that the RTT changes is caused by the path change. We say that these two changes are correlated.

However, there is no straightforward matching between the two changes, as the measurement intervals are different: $30min$ for traceroute while $4min$ for ping. 
Again, \textit{minimum cost maximum-cardinality matching} appears to be a reasonable formulation of the correlation between RTT and path changes.
We therefore borrow the concept of optimal matching in changepoint evaluation (Sec.~\ref{sec:score}).
The shift tolerance window is set to the interval of traceroute measurement, as causal relationship between the RTT and path change is possible within that range.
A pair of RTT and path changes are correlated/matched if they are within in the so produced optimal matching.
The notion $Precision$, introduced in section~\ref{sec:score}, is now interpreted as the fraction of path changes that are matched to an RTT change. 

\begin{figure}[!htb]
\centering
\includegraphics[width=.36\textwidth]{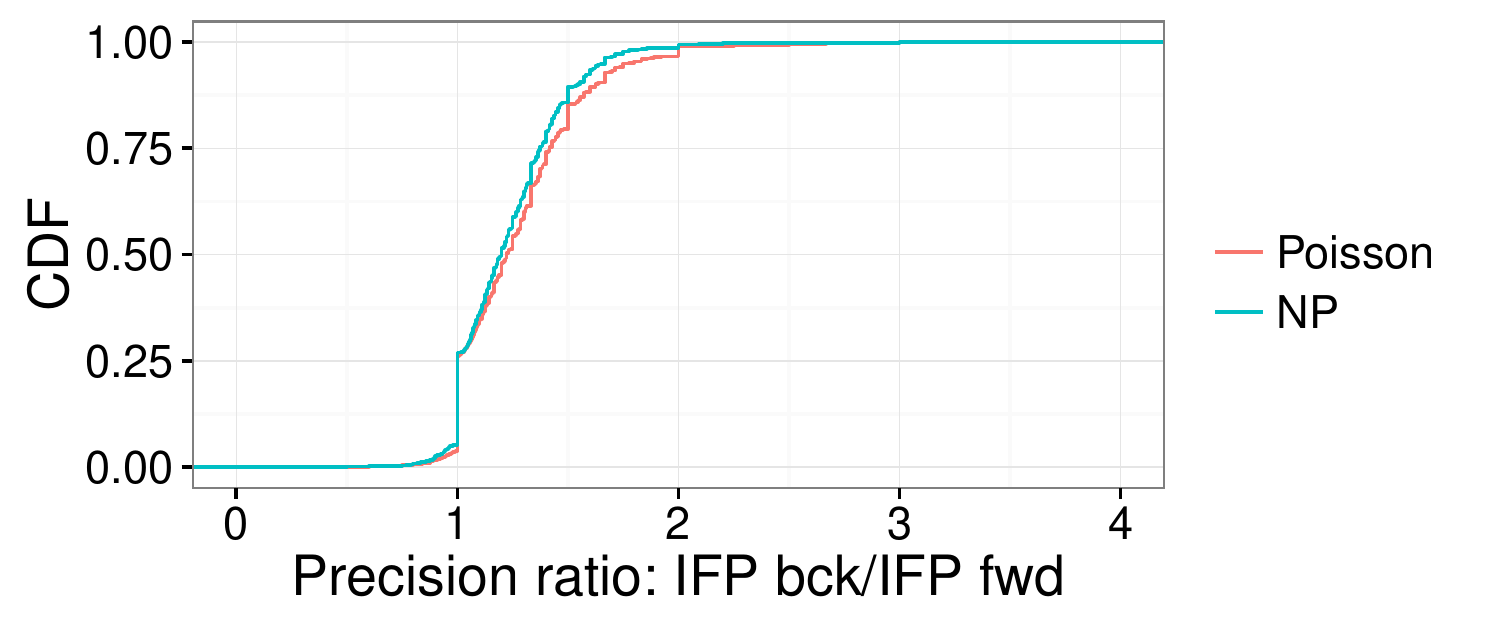}
\caption{Precision ration between IFP changes detected by \textit{backward extension} and \textit{forward inclusion}.}
\label{fig:ifp_bck_ch_precision_gain_cdf}
\end{figure}
Fig.~\ref{fig:ifp_bck_ch_precision_gain_cdf} shows that with same number of changes, IFP changes detected with \textit{backward extension} are much more likely to have a match with RTT changes for $75\%$ probes.
The significant increase in precision implies the occurrences of short IFP deviations, as well confirms the success of \textit{backward extension} in capturing them.

\begin{table}[!htb]
\caption{Number of RTT changes matched with a path change for the selected 2050 probes.}
\label{tab:corr_overview}
\centering
\footnotesize
\setlength{\tabcolsep}{0.5em}
\begin{tabular}{l|cc|c}
\toprule
& \texttt{cpt\_poisson} & \texttt{cpt\_np} & \# path changes\\
\midrule
AS path change & 11,794 & 6,380 & 51,282 \\
IXP change & 9,126 & 8,341 & 73,544\\
IFP change & 38,700 & 36,400 & 244,713\\
\midrule
\# RTT changes & 481,877 & 307,312 & \\
\bottomrule
\end{tabular}
\end{table}
Tab.~\ref{tab:corr_overview} details the number of matched RTT and path change pairs.
A large fraction of path changes doesn't match with any RTT changes.
Especially, the fraction of AS path changes matched to RTT changes with either detection method is much lower than the reported $72.5\%$ in \cite{Rimondini2014}.
An important part of RTT changes detected by both method doesn't match with any change in the forwarding path either.
Moreover, the number of RTT changes correlates with AS path changes differs greatly across the two changepoint methods,
while the number matched to IXP or IFP change are relatively close.
We explore the underlying reasons behind above phenomena.

\subsection{\texttt{cpt\_poisson} matches better with AS path change?}
\label{sec:as_match_diff}
\begin{figure}[!htb]
\centering
\includegraphics[width=.48\textwidth]{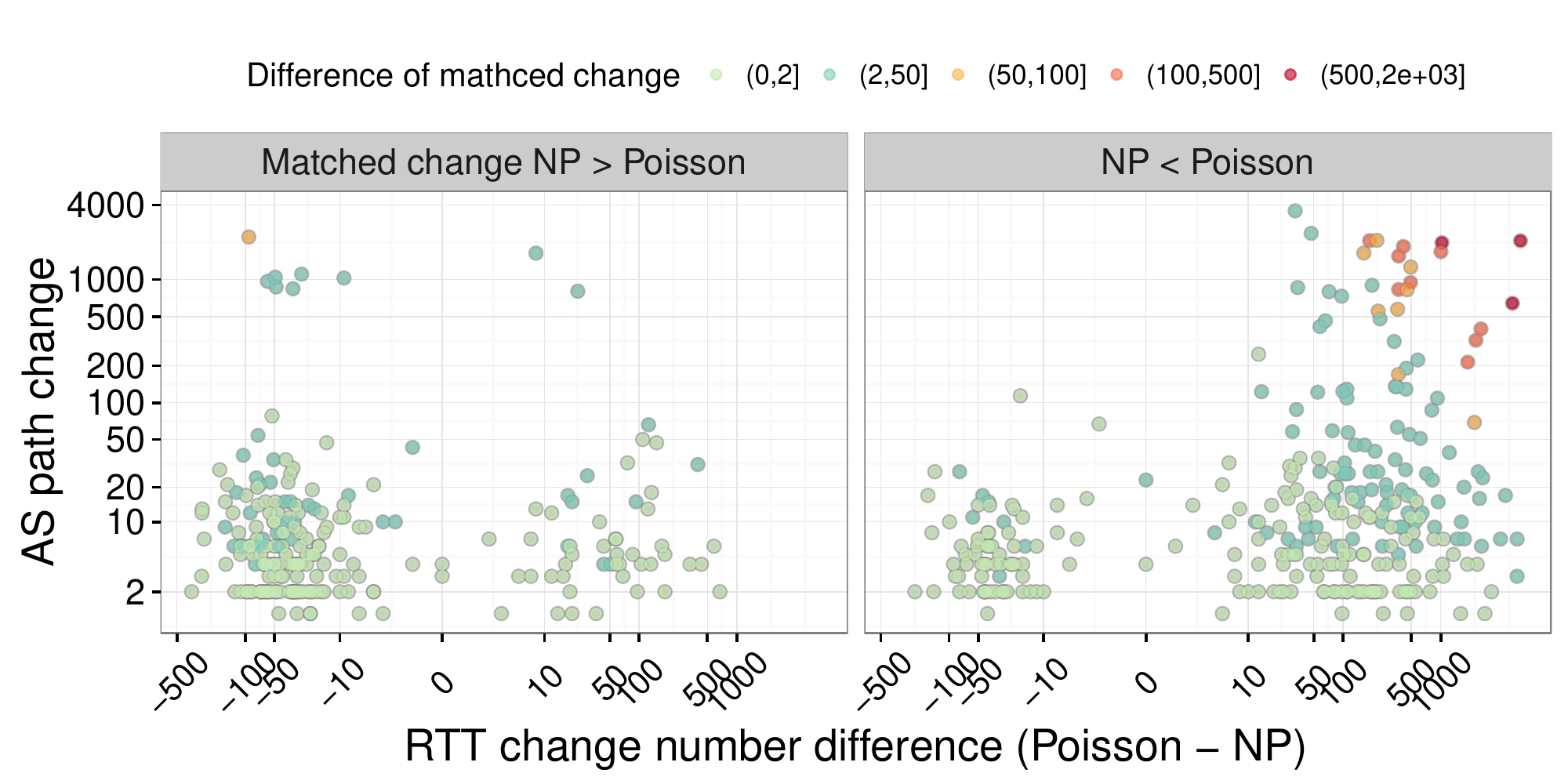}
\caption{Probe having difference in the number of AS path changes matched to RTT changes detected by \texttt{cpt\_poisson} and \texttt{cpt\_np}. Probes are characterized by its AS path change numbers and RTT change number difference between the two methods. The color of each probe indicates the level of difference in matched change. Left panel shows the probes with more AS path changes matched to \texttt{cpt\_np} RTT changes.}
\label{fig:as_match_diff}
\end{figure}

Among 880 probes ever experienced AS path changes, 293 probes have more AS path changes matched to \texttt{cpt\_poisson} RTT changes, 224 have more AS path changes matched to \texttt{cpt\_np} changes, 
Among these 517 probes, 463 are with a difference smaller than 10 changes.
The rest 363 probes have no difference across the two methods.
Contrary to what we see in Tab.~\ref{tab:corr_overview}, the numbers of AS path matched are in fact highly consistent across the two methods for the majority of probes.
The difference is caused by a small fraction of probes identified in Fig.~\ref{fig:as_match_diff}
More the color (of a probe) is on the red side, more important the difference is between \texttt{cpt\_poisson} and \texttt{cpt\_np}.
We can tell that those probes plainly in red all experienced a large number of AS path changes (y-axis).
Moreover, if more path changes are matched to RTT changes detected by one method, very likely more RTT changes are as well detected by this method than the other.
All together, the difference in matched RTT changes with AS path changes fundamentally lies in the difference of detection sensitivity (number of detected RTT changes) across different probe traces. This difference is manifested through extremely frequent AS Path changes of several specific probes.

\begin{figure}[!htb]
\centering
\includegraphics[width=.32\textwidth]{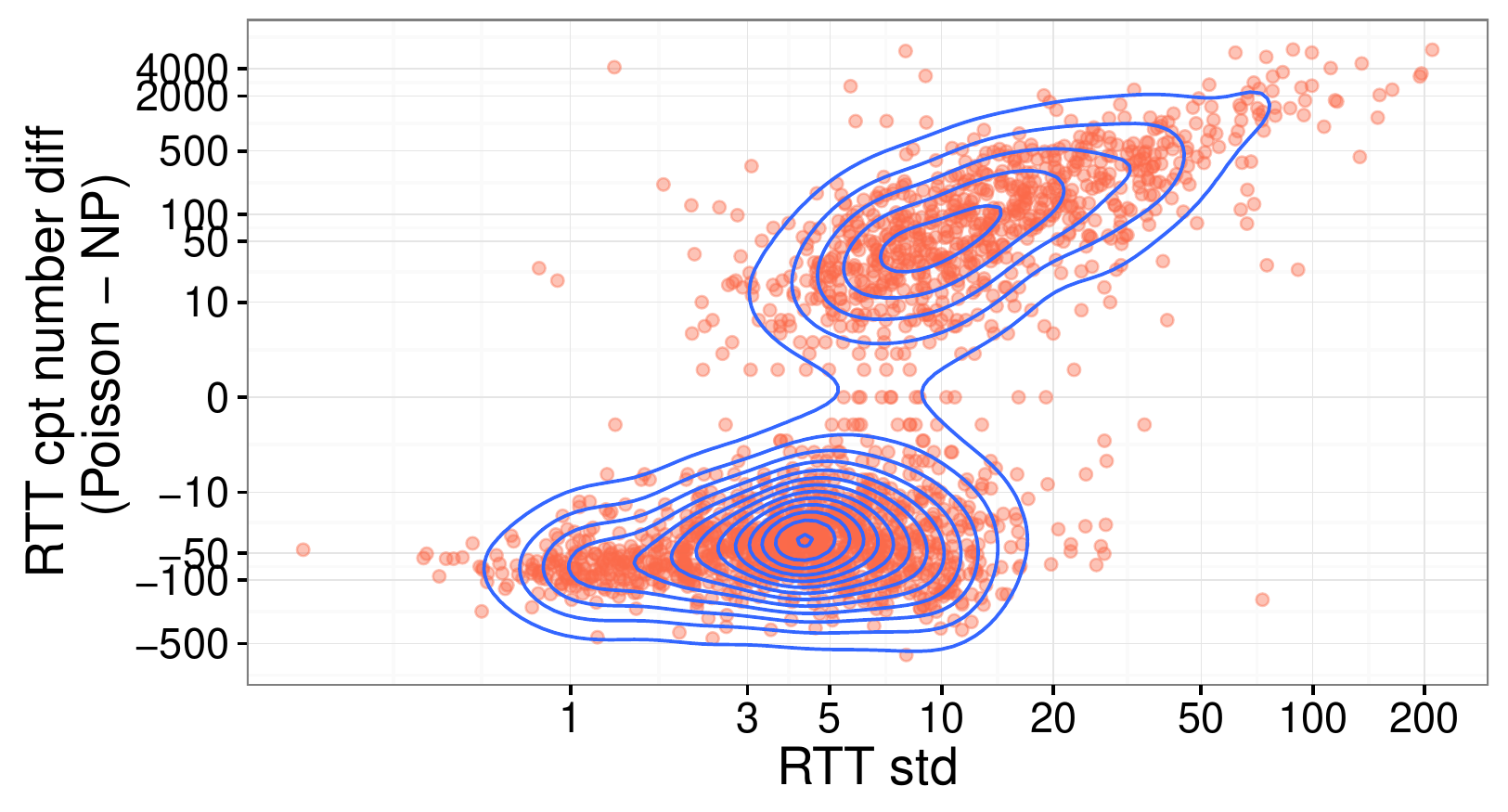}
\caption{Relation between RTT change number difference by the two changepoint method and the RTT trace $std$.}
\label{fig:cpt_diff_vs_std}
\end{figure}
Fig.~\ref{fig:rtt_ch_count_density_cmp} and Tab.~\ref{tab:corr_overview} already tell that \texttt{cpt\_poisson} detects in total more RTT changes than \texttt{cpt\_np}, appearing to be more sensitive.
However, for most probes with small overall RTT variation, \texttt{cpt\_np} is in fact more sensitive and detects more RTT changes, according to Fig.~\ref{fig:cpt_diff_vs_std}.
This matches with Fig.~\ref{fig:rtt_ch_count_density_cmp} in the sense that \texttt{cpt\_np} detects much more changes of small amplitude.
For probes with relatively large overall RTT variation, \texttt{cpt\_poisson} tends to be more sensitive and the difference in change number increases with the level of RTT variance.
With comprehensive manual inspection, we found that those RTT traces with high variance mostly underwent large amplitude RTT oscillations, many of which caused by ping timeouts. As human change detector, we also found very difficult to mark moments of change for these traces.

\subsection{How AS path changes match to RTT changes?}
\begin{figure}[!htb]
    \centering
    \begin{subfigure}[b]{.24\textwidth}
	\centering
	\includegraphics[width=\textwidth]{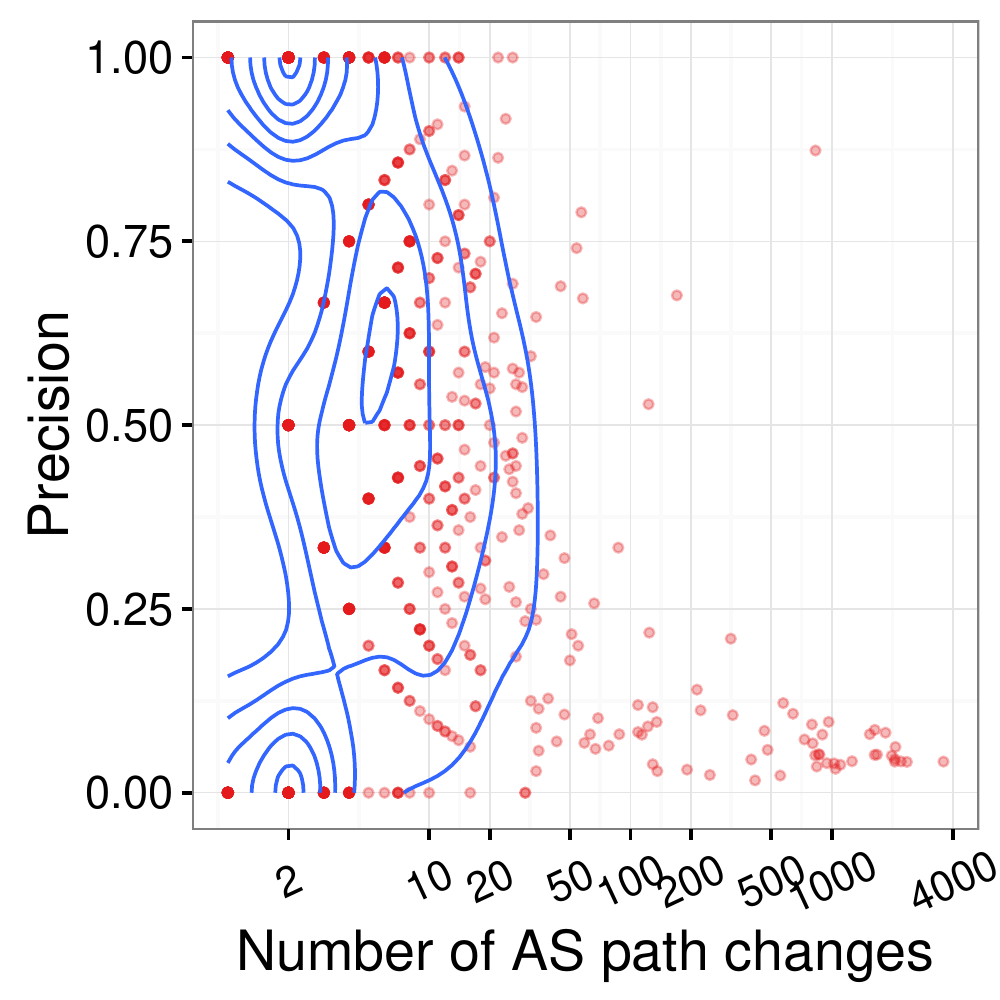}
	\caption{\footnotesize changes by \texttt{cpt\_np}.}
	\label{fig:as_path_ch_precision_np}
	\end{subfigure}
	\begin{subfigure}[b]{.24\textwidth}
	\centering
    \includegraphics[width=\textwidth]{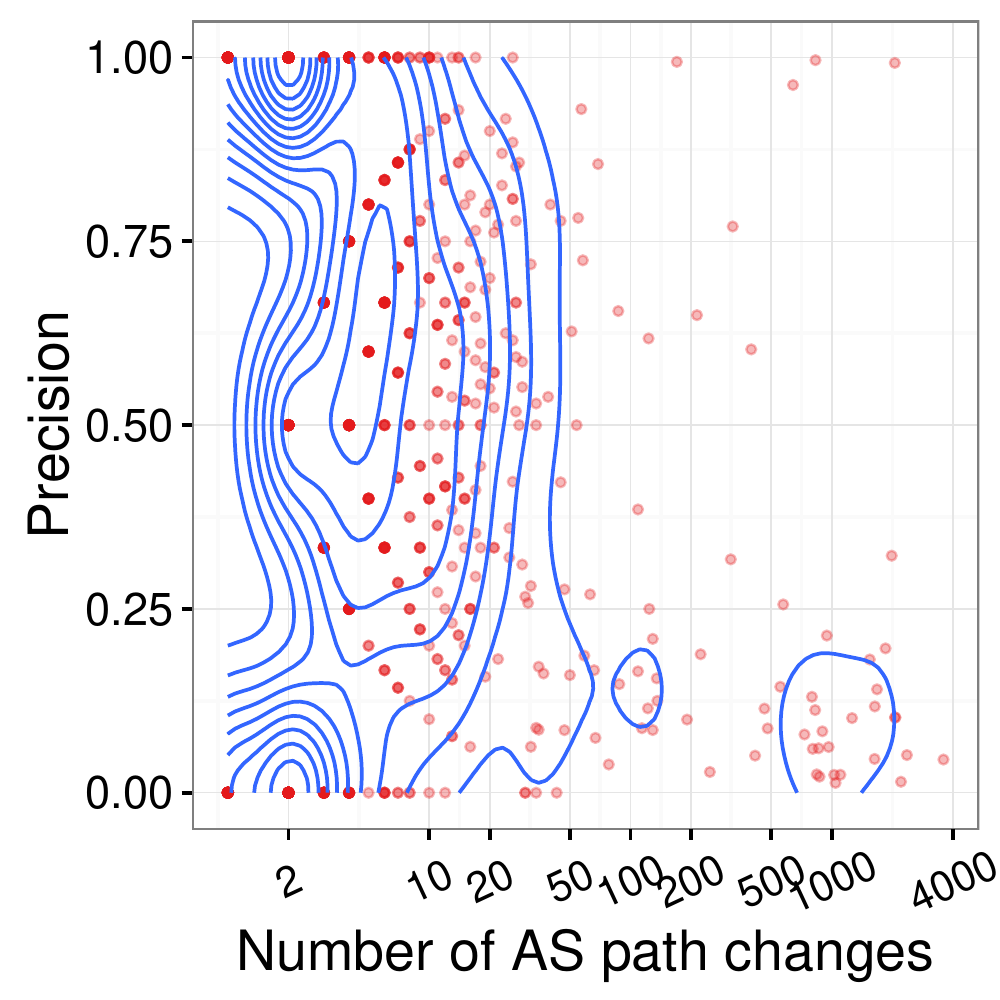}
	\caption{\footnotesize changes by \texttt{cpt\_poisson}.}
	\label{fig:as_path_ch_precision_poisson}
	\end{subfigure}
\caption{The relation between precision and AS path change times per probe trace.}
\label{fig:as_path_ch_precision}
\end{figure}
\begin{figure}[!htb]
\centering
\includegraphics[width=.48\textwidth]{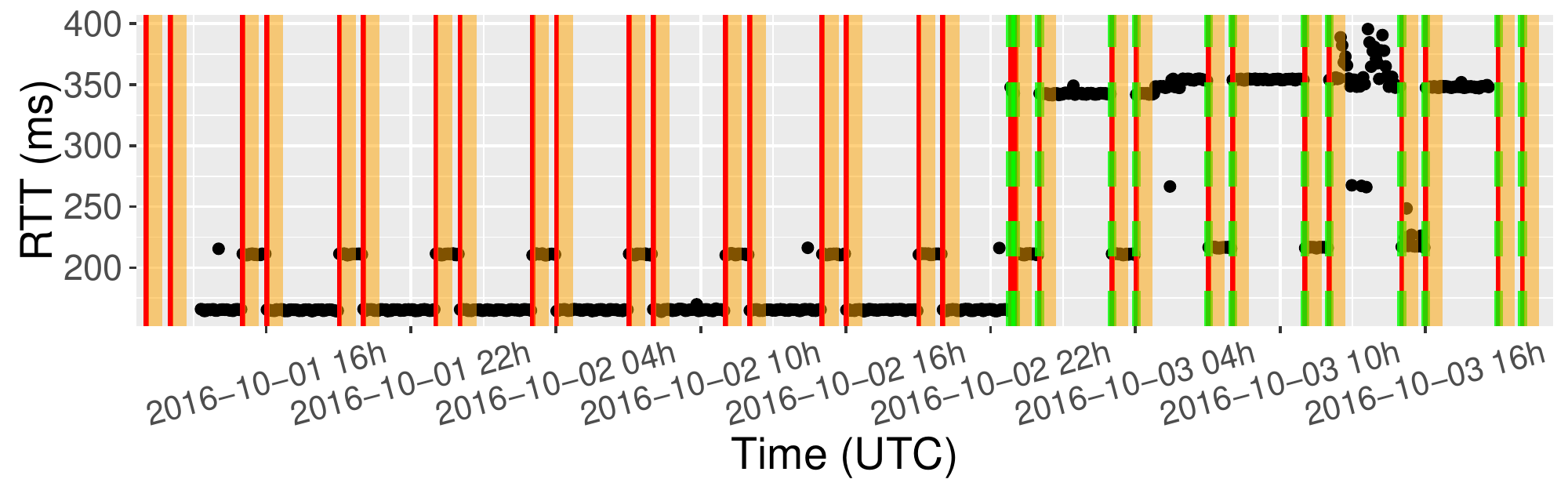}
\caption{RTT from Probe 12849. Red lines for RTT change detected by \texttt{cpt\_poisson}; green dotted lines for RTT change by \texttt{cpt\_np}. Orange strips for AS path changes.}
\label{fig:case_12849_rtt}
\end{figure}
A nature question following the above analysis is whether those frequent AS path changes can in fact cause RTT changes.
Fig.~\ref{fig:as_path_ch_precision} reveals that probes with extremely frequent AS path changes have in fact very low correlation, in terms of precision, with RTT changes.
After extensive manual inspection, those frequent AS path changes appear to be AS-level LB, i.e. upstream ASes in reaching the destination are switched very frequently among a few providers.
Such AS changes generally don't have a clear impact on the RTT level.

Yet, not all such AS-level load balancing is without consequence. For example, probe 12849 in Fig.~\ref{fig:case_12849_rtt} experienced 170 AS path changes, among which 169 are matched to RTT changes detected by \texttt{cpt\_poisson} and only 102 are matched to \texttt{cpt\_np}.
These AS path changes are highly periodic and coincide with clear cut RTT changes.
\texttt{cpt\_np} failed to detect some of the changes with smaller amplitude.

\subsection{Pitfalls of IXP and IFP change detection}
Similar to AS path changes, the correlation with RTT changes is weak for frequent IXP and IFP changes.
In Fig.~\ref{fig:path_ch_count_density_cmp}, a group of probes are in the area of 100 to 200 IXP changes. Their correlation with RTT changes is fairly low, precision  around $0.1$.
We investigate all the 58 probes in the area. These probe passed by AMS-IX to reach b-root most of the time. There were about 147 times, shared by these probes, where AMS-IX hop was replaced by a timeout hop before arriving AS6939.
In such case, no IXP related address appears in the measured path. The presence of IXP is thus uncertain \footnote{The newly released traIXroute v2.1 can detect IXP without the presence of IXP related IP address, if the neighbouring ASes are known to be member of a same IXP. However, it is still possible that two ASes peer at multiple IXPs, where the exact IXP traversed would remain uncertain.}.

\begin{table}[!htb]
\caption{Quantiles of IP path numbers per probe trace.}
\label{tab:ip_path_count}
\centering
\footnotesize
\setlength{\tabcolsep}{0.5em}
\begin{tabular}{ccccccc}
\toprule
$5\%$ & $10\%$ & $25\%$ & $50\%$ & $75\%$ & $95\%$ & $100\%$\\
\midrule
20 & 32 & 56 & 91 & 145 & 419 & 4302\\
\bottomrule
\end{tabular}
\end{table}
The correlation of IFP changes with RTT changes are much weaker than that of AS and IXP path changes.
It turned out that most probes experienced much more than 16 end-to-end IP paths, Tab.~\ref{tab:ip_path_count}.
In such case, one Paris ID might have been mapped to more than one IP path in reality, which leads to IFP changes without actual routing change in the forward path.
Within in each single AS, the number of different IP paths rarely exceeds 16 toward a destination. However a chain of ASes can produce way much richer combinations of end-to-end IP paths.

Moreover, there is a group of probes having around 250 IFP changes according to Fig.~\ref{fig:path_ch_count_density_cmp}.
An IFP change takes place roughly every 16 measurements on these probes. 
These changes are as well poorly correlated to RTT changes.
We investigated some probes in the area and found out the frequent changes aren't necessary related to the large amount of end-to-end paths.
For some probes, two neighbouring IFPs only differ at one or two Paris IDs.
IP paths taken by these Paris IDs oscillates between a few alternatives frequently.
For example, the Paris ID 6, 7, 8, 9 of probe 23998 switches a lot among only 2 paths.
Such change in general doesn't have obvious consequence on RTT level.

\subsection{Unmatched RTT changes}
\begin{figure}[!htb]
    \centering
    \begin{subfigure}[b]{.48\textwidth}
	\centering
	\includegraphics[width=\textwidth]{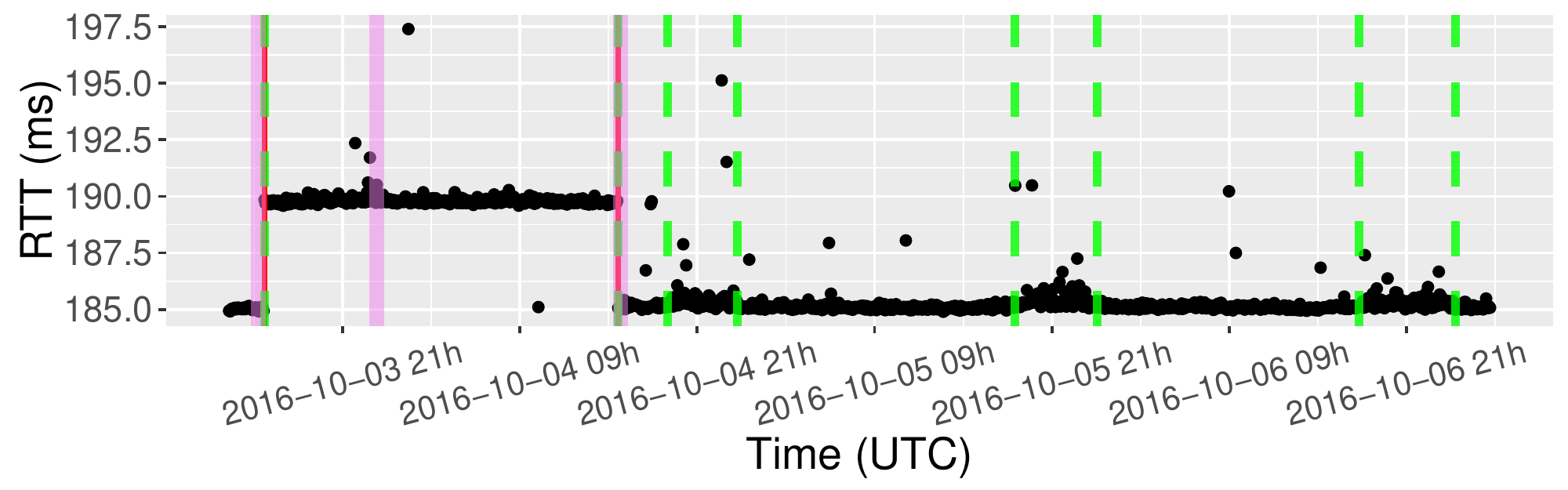}
	\caption{\footnotesize Probe 28002.}
	\label{fig:case_28002}
	\end{subfigure}
	\begin{subfigure}[b]{.48\textwidth}
	\centering
	\includegraphics[width=\textwidth]{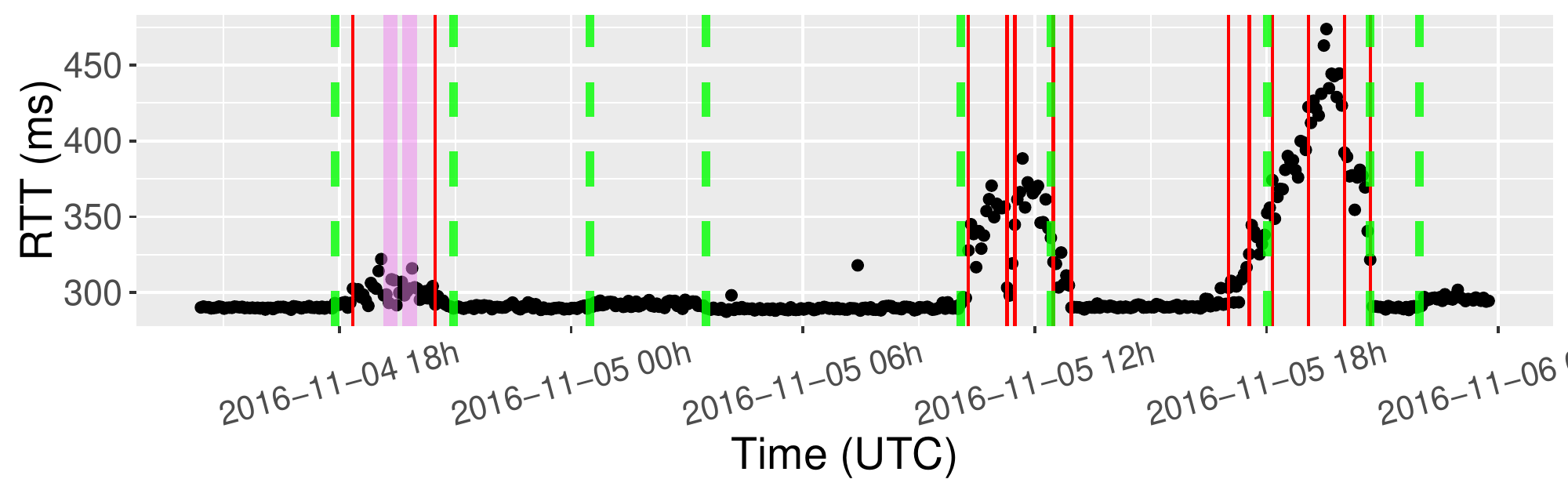}
	\caption{\footnotesize Probe 26328.}
	\label{fig:case_26328}
	\end{subfigure}
\caption{RTT trace and change detection example. Red lines for RTT change detected by \texttt{cpt\_poisson}; green dotted lines for RTT change by \texttt{cpt\_np}. Violet strips are IFP changes.}
\label{fig:case_sensitivity}
\end{figure}

Several reasons contribute to the large amount and fraction of RTT changes unmatched to any path changes.
First, some path changes experienced by RTT measurements are not observed. We were not able to measure the reverse path with RIPE Atlas built-in measurement, let alone detecting the changes on the reverse path. However, these changes could have contributed to RTT changes, especially in the context of inter-domain routing where paths are likely to be asymmetric.
Second, congestion. Fig.~\ref{fig:case_26328} gives an typical example of RTT changes probably caused by congestion.
Congestion like this doesn't repeat periodically, and thus can not be detected with existing methods~\cite{Luckie2014}.
Changepoint methods studied in this work can be potentially employed to pinpoint such transient congestion and estimate their
impact on the tranmission performance. We envision it as future work.
Third, RTT change detection can be over-sensitive. We revealed from a macroscopic view in Sec.~\ref{sec:cpt_trace} and ~\ref{sec:as_match_diff} that \texttt{cpt\_poisson} tend to overestimate the number of changepoints when the RTT trace is noisy, while \texttt{cpt\_np} is capable of detecting delicate RTT changes. Individual traces are given in Fig.~\ref{fig:case_sensitivity} to illustrate the sensitivity difference from a microscopic view. In Fig.~\ref{fig:case_28002}, \texttt{cpt\_np} detected all the periodic small amplitude congestion. In Fig.~\ref{fig:case_26328}, both methods identified the two large `plumbs' near the end of the trace. The difference is that \texttt{cpt\_poisson} marked intermediate level changes as well.

\section{Conclusion}
\label{sec:cls}
In this paper, we proposed an evaluation framework for change detection on RTT time series.
The framework is robust with human-labelled dataset and weights RTT changes according to their importance in network operation. 
In detecting path changes, we distinguish those caused by routing changes from those due to load balancing.
Finally, we correlate the detected RTT and path changes by establishing an one-to-one matching between them. 
We investigated the sensitivity distinction across different change detection methods. 
Hidden issues with path changes are as well revealed.

This work is mere a facilitator for measurement-based TE. Further efforts are required in building a working system. To name a few: online detection of RTT changes, route selection logic triggered by change detection  etc.

\bibliographystyle{IEEEtran} 
\bibliography{IEEEabrv,ref}
\end{document}